\documentclass[showpacs,aps,pre,twocolumn]{revtex4}
\usepackage{graphicx}
\usepackage{amsfonts}

\begin{document}
\title{Soliton dynamics at an interface between uniform medium and nonlinear optical lattice}

\author{Fatkhulla Kh. Abdullaev$^{1}$,
\footnote[7]{Corresponding author (fatkhulla@yahoo.com)} \, Ravil
M. Galimzyanov$^{1}$,\, Marijana Brtka$^{2}$, \, and Lauro
Tomio$^{3}$} \affiliation{$^1$ Physical-Technical Institute of the
Academy of Sciences, G.Mavlyanov str.,2-b, 100084, Tashkent, Uzbekistan\\
$^2$ Instituto de F\'isica, Universidade de S\~ao Paulo,
05315-970, C.P. 66318, S\~ao Paulo, SP, Brazil \\
$^3$ Instituto de F\'\i sica Te\'orica,
S\~ao Paulo State University (UNESP),\\
Rua Pamplona, 145, 01405-900, S\~ao Paulo, Brazil }

\begin{abstract}
We study trapping and propagation of a matter-wave soliton through
the interface between uniform medium and a nonlinear optical
lattice (NOL). Different regimes for transmission of a broad and a
narrow soliton are investigated. Reflections and transmissions of
solitons are predicted as function of the lattice phase. The
existence of a threshold in the amplitude of the nonlinear optical
lattice, separating the transmission and reflection regimes, is
verified. The localized nonlinear surface state, corresponding to
the soliton trapped by the interface, is found. Variational
approach predictions are confirmed by numerical simulations for
the original Gross-Pitaevskii equation with nonlinear periodic
potentials.
\end{abstract}
\pacs{
03.75.Lm, 
05.45.Yv, 
02.30.Jr, 
42.65.Tg  
}
\maketitle

\section{Introduction}
Investigation of processes of reflection, transmission and
trapping of a nonlinear wave packet at the interface between two
different nonlinear media represent one of the fundamental
problems of the nonlinear physics~\cite{ANM,KKC}. Recently, the
problem of reflection and/or transmission of a soliton at the
interface between a nonlinear uniform media and a linear periodic
structure, under the conditions of the Bragg resonance, has been
considered in Ref.~\cite{Kartashov06}. There it was shown the
possibility of controlling such a structure in the regime of the
soliton mirror.

Considering the actual experimental possibilities, it will be also
of interest to study the transmission and trapping phenomena when
we have a periodic variation in space of the parameter related to
the nonlinearity. Such system can be realized in Bose-Einstein
condensates (BEC) by using a periodically modulated in space
external magnetic field or optically induced Feshbach
resonances~\cite{Kagan,NOL1,NOL2}. Standing optical wave can
induce in BEC periodic space modulation in the atom-atom
scattering length. In the Gross-Pitaevskii (GP) equation it leads
to periodic space modulations of the mean field nonlinearity;
i.e., producing a
NOL~\cite{AS03,SM06,AG05,kevrekidis,Garcia07,Abd08,Kart,Blud,Konotop}.
Considering the two-component case in the 1D limit, the properties
of BEC confined in NOL, as well as existence of soliton solutions
and their stability, are investigated in Ref.~\cite{AGST}. BEC
with finite segment of periodically space-modulated atomic
scattering length (shallow optical lattice) are considered in
Ref.~\cite{Dong}, where matter-wave optical limits and bistability
are predicted. For the dynamics of matter wave propagation under
different conditions, see the review~\cite{ijmpe} and references
therein. Gap solitons are analyzed in Ref.~\cite{AAG}, where it
was shown that localized nonlinear wave packets can exist in NOL
for attractive condensates (bright solitons) as well as for
repulsive ones (dark solitons). The stability analysis showed that
the bright solitons are stable in a very narrow region of
parameters~\cite{fibich}. However, such analysis is absent in case
of the existence of an interface. Linear surface states in
lattices with management of the diffraction have been recently
considered in Ref.~\cite{Garanovich}. Surface soliton formation at
an interface between two periodic media is studied in
Ref.~\cite{Kominis}.

In the present paper we consider regimes of reflection,
transmission and trapping of a matter wave soliton incident on the
interface between uniform medium and a nonlinear optical lattice.
Particular attention will be devoted to the possible existence of
nonlinear surface states for matter waves. An interface induces
changes in the effective potentials for the soliton center and
width and can create a surface soliton. Also the stability can be
enhanced. The dynamics of a BEC in a quasi-one-dimensional
elongated trap will be treated by considering the GP formalism
reduced to the one-dimensional (1D) space limit.

Recently, an investigation done in Ref.~\cite{Kartashov3}
considered two-dimensional nonlinear surface states (surface
solitons) at an interface in a superposition of a periodic
potential and periodic modulations of the nonlinear space
parameter. The physical system is motivated by optical structures
writing on quartz by femtosecond laser (fs-laser). In this case,
variations of the Kerr nonlinearity remain of the same sign and
are out of the phase with the periodic variations of the linear
refractive index.
As opposed to this nonlinear optical system, in BEC case we can
also realize the cases of periodic modulations when the
nonlinearity changes sign.

The transmission characteristics of solitons are defined by the
effective potential induced by the interface and nonlinear
periodic lattice. The effective potential strongly depends on the
soliton parameters. Unlike the soliton transmission
through the linear lattice, in the case of nonlinear periodic
lattice we have nontrivial intensity-dependence (number of atoms)
for the form of the potential relief as well as a {\it threshold}
behavior depending on the amplitude of the nonlinearity
modulations in space. It means that  by change in the amplitude of
modulations performed by variation of external magnetic field near
the Feshbach resonance point, we can form a mirror for the
matter-wave solitons, selecting the solitons by the number of
atoms.

The paper is organized as follows. In section 2 the model is
formulated and variational equations for the soliton parameters
are derived. In section 3 characteristics of stationary soliton
trapped by interface are investigated. The reflection and
transmission regimes for narrow and broad solitons are analyzed.
The summary of obtained results is given in conclusion.

\section {The model}
In order to describe the propagation of a matter wave soliton in
the elongated quasi-1D condensate with attractive interaction, we
consider the GP equation in a 1D space approach where the physical
space-time variables are given by $(\overline{x},\overline{t})$, and the
corresponding dimensionless variables are $(x,t)$:
\begin{equation}
{\rm{i}}\hbar\frac{\partial\psi}{\partial{\overline{t}}} +
\frac{\hbar^2}{2m} \frac{\partial^2\psi}{\partial{\overline{x}^2}}  +
g_{1D}(\overline{x})|\psi|^2 \psi =0,
\label{GP1}
\end{equation}
where $\psi\equiv\psi(\overline{x},\overline{t})$ and $g_{1D}(\overline{x})
\equiv 2\hbar a_s(\overline{x})\omega_{\perp}.$ Here, $\omega_{\perp}$ is
the transverse frequency of the trap and $a_s(\overline{x})$ the
spatially dependent atomic scattering length, which is supposed to
vary in space for $\overline{x}>0$ as $a_s(\overline{x})= a_0 +
\theta(\overline{x})\left(\delta_0 + a_{1}\sin(2 k \overline{x})\right)$.
Here, $\theta(x)=0 (1)$ for $x<0 (x>0)$, $\delta_0$ is a constant,
$a_0$ is the natural two-body scattering length, and the wave
number $k$ is related to the lattice period $L$ by $k\equiv
2\pi/L$. The number of atoms $\overline{N}$ normalizes the
wave-function as
\begin{equation}
\overline{N} = \int_{-\infty}^{\infty} |\psi|^2 d\overline{x}.\label{eq2}
\end{equation}
To avoid the collapse in the attractive BEC, the condition
$|a_s|\overline{N}/\omega_{\perp} < 0.676$ should be
satisfied~\cite{gtf}.

The transformation to the new set of dimensionless space-time
variables $(x,t)$ is given by the following:
\begin{eqnarray*}
x \equiv k \overline{x}, \;\; t \equiv \omega_R \overline{t},\;\;
\gamma(x)\equiv \displaystyle \frac{a_s(\overline{x})}{|a_0|},\;\;
\Delta_0\equiv\frac{\delta_0}{|a_0|},\\
E_R \equiv \hbar \omega_R  \equiv \displaystyle \frac{\hbar^2 k^2}{2m},\;\;
 u\equiv u(x,t) \equiv
 \sqrt{\frac{2\omega_{\perp}}{\omega_R}|a_0|}\psi.
\label{transf}
\end{eqnarray*}
With the above, where $E_R$ is the recoil energy, we obtain the
dimensionless form of the 1D GP equation:
\begin{eqnarray}
&&{\rm i}u_t + u_{xx}  + \gamma(x)|u|^2 u = 0,\label{eq1} \\
&&\gamma(x) = [\gamma_{0}+\theta(x)(\Delta_0 + \gamma_{1}\sin(2
x))],\nonumber
\end{eqnarray}
where $\gamma_0 = a_0/|a_0|=\pm 1$ (for the attractive and
repulsive condensates respectively) and  $\gamma_1 = a_1/|a_0|$.
In the above equation and in the following, we use the abbreviated
notation for partial differential equations, such that
$u_t\equiv {\partial{u}}/{\partial t}$.
The normalization of
$u$, $N$, relates to the number of atoms $\overline{ N }$, which is
conserved. From Eqs.~(\ref{eq2}) and (\ref{transf}), we obtain:
\begin{equation}
{N} = \int_{-\infty}^{\infty} |u|^2 dx =
\frac{4m|a_0|\omega_\perp}{\hbar k}\overline{N}.\label{norm2}
\end{equation}
Below we will consider the evolution of bright solitons ($\gamma_0
= 1$). When solitons collide at the interface ($x=0$), different
scenarios are possible resulting in reflection, transmission or
trapping. Let us consider different limiting cases of broad and
narrow solitons (with respect to the period of modulations). To
study the soliton evolution we shall use the variational
approach~\cite{anderson}. According to this method we should
calculate an averaged Lagrangian and then, using the
Euler-Lagrange equations, obtain the equations for the soliton
parameters.

\begin{figure}[htbp]
\begin{center}
\includegraphics[width=8cm,height=6cm,angle=0]{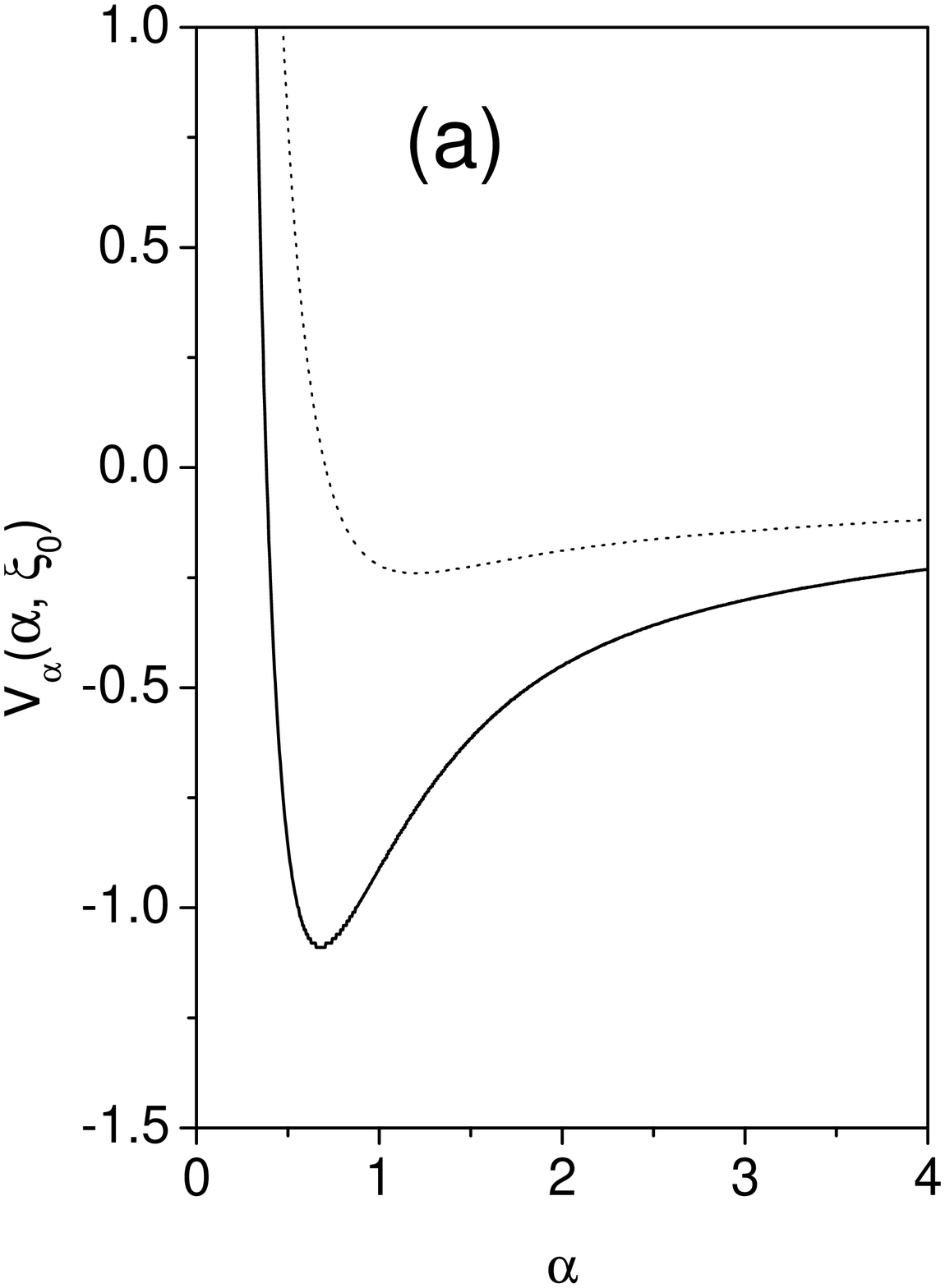}
\includegraphics[width=8cm,height=6cm,angle=0]{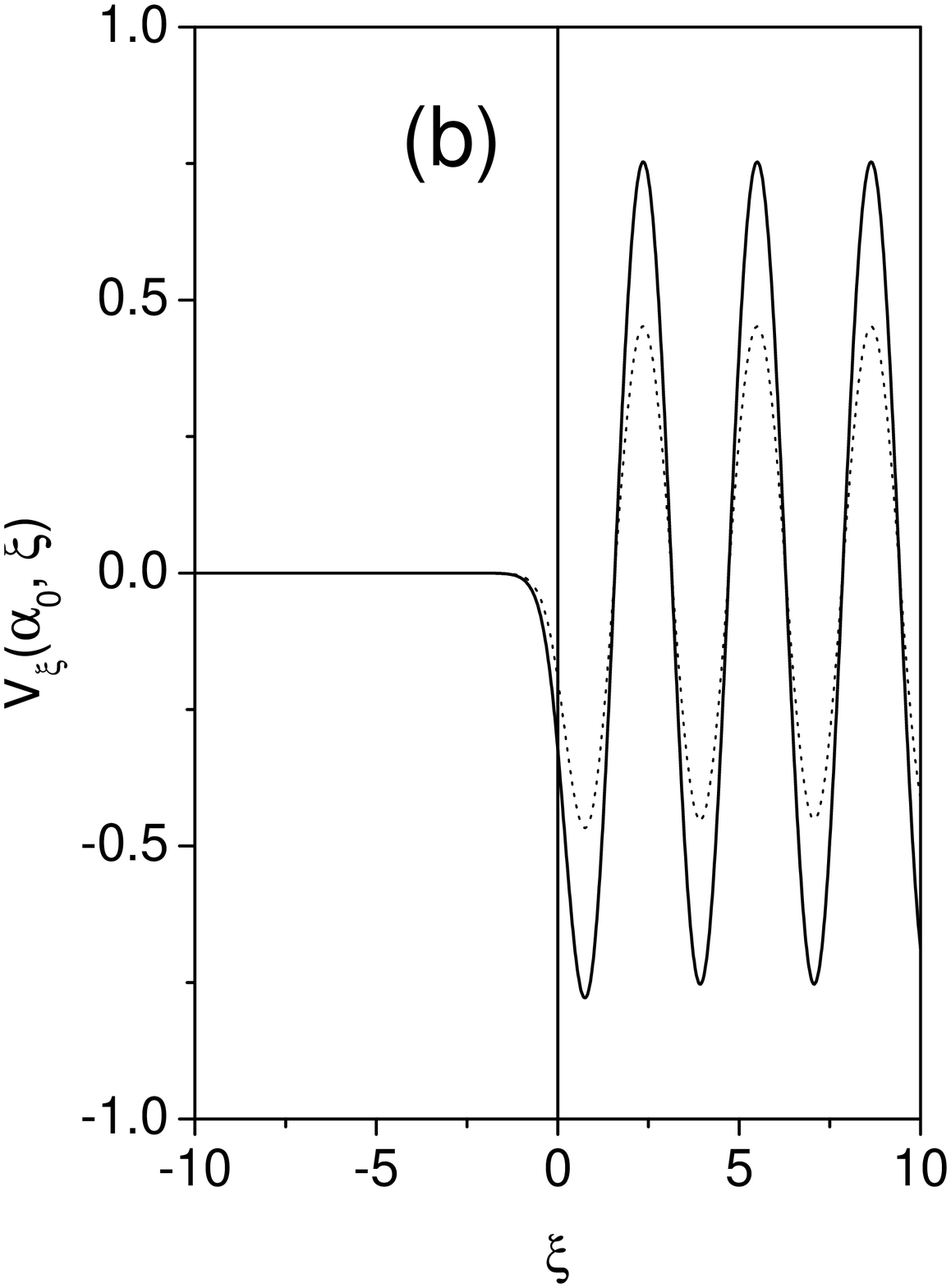}
\end{center}
\caption{ Profiles of the effective potentials
$V_\alpha(\alpha,\xi_0)$ [as a function of the width $\alpha$,
shown in (a)] and $V_\xi(\alpha_0,\xi)$ [as a function of $\xi$,
shown in (b)], where values of $\alpha_0$ and $\xi_0$ correspond
to the stationary point of the set of equations (\ref{Vareq_x}),
(\ref{Vareq_a}). Vertical line in (b) indicates the {\it
interface}. Two cases are depicted, $N = 1.5$ with
$\alpha_0=1.213$, $\xi_0=0.668$ (dotted line) and $N = 2.5$ with
$\alpha_0=0.674$, $\xi_0=0.773$ (solid line). Other parameters are
$\gamma_0=1$, $\gamma_1=1.2$, $\Delta_0=0.$ } \label{Veff}
\end{figure}

\begin{figure}[htbp]
\vspace{0.5cm}
\begin{center}
\includegraphics[width=8cm,height=6cm,angle=0]{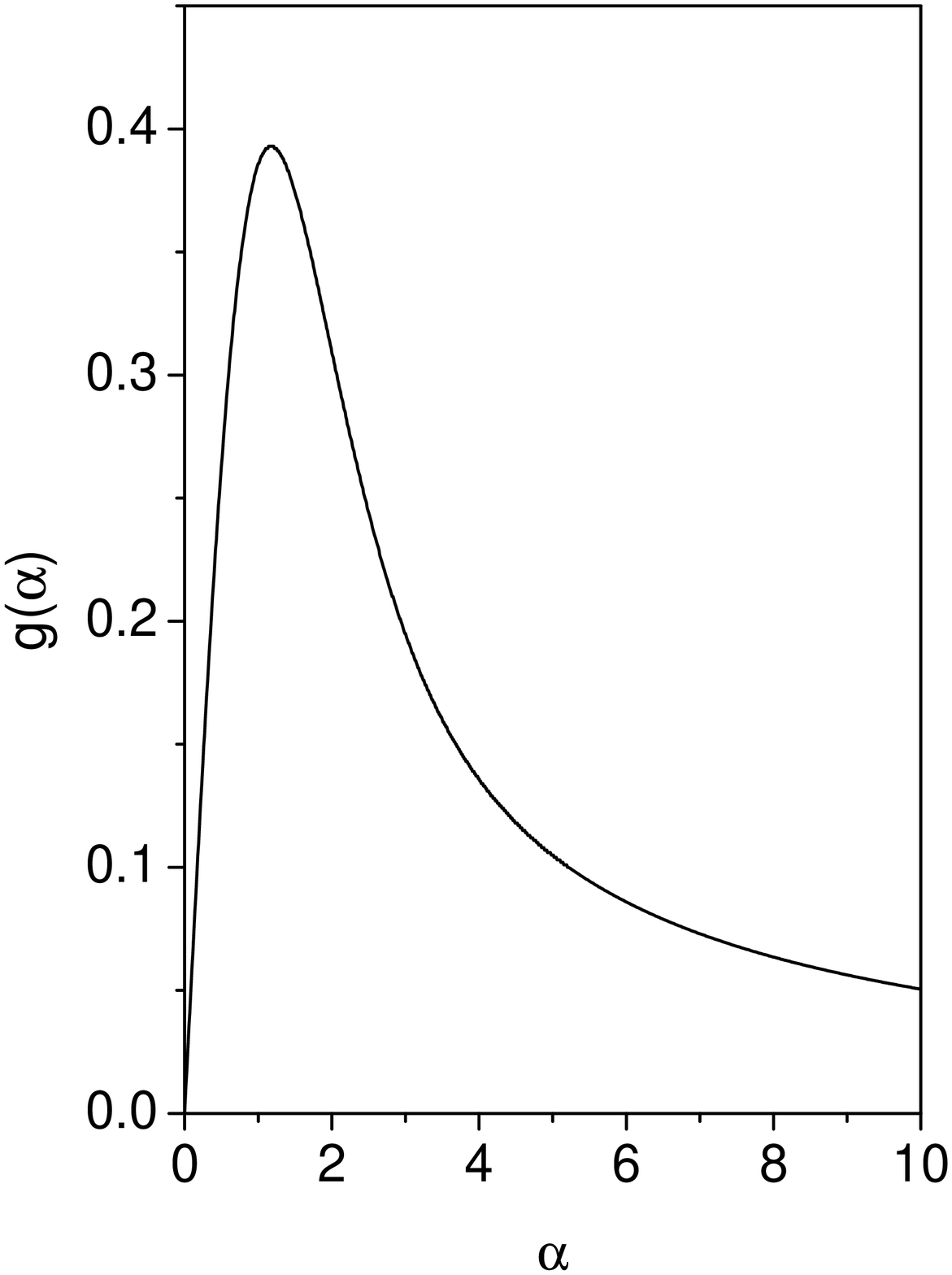}
\end{center}
\caption{Factor $g(\alpha)$ versus the soliton width $\alpha$.}
\label{g}
\end{figure}

The Lagrangian density corresponding to Eq.~(\ref{eq1}) is given by:
\begin{equation}
L = \frac{i}{2}(u_t u^{\star} - u^{\star}_t u)-|u_x|^2  +
\frac{1}{2}\gamma(x)|u|^4.
\end{equation}
In deriving our variational model we proceed from the following
anzatz for a soliton:
\begin{equation}\label{trf}
u = \sqrt{2}A \mbox{sech}\left(\frac{x-\xi}{\alpha}\right)
 e^{ {\rm i}\beta(x-\xi)^2 + {\rm i}\kappa(x-\xi)+ {\rm i}\phi}.
\end{equation}
In order to obtain the equations for the soliton parameters
($A,\alpha,\beta,\kappa,\xi,\phi$), we
calculate the averaged Lagrangian
$\overline{L}=\int_{-\infty}^{\infty}L(x,t)dx$ with the above trial function
(\ref{trf}):
\begin{eqnarray}
\frac{\overline{L}}{N} = -\frac{\pi^2}{12}\beta_t \alpha^2+\kappa\xi_t
-\phi_t -
\frac{1}{3\alpha^2} -\frac{\pi^2}{3}\beta^2\alpha^2-\kappa^2 +\nonumber\\
  \frac{N\gamma_0}{6\alpha} + \frac{\Delta_0 N}{8\alpha}F_0(\xi,\alpha) +
\frac{\gamma_1 N}{8\alpha}F_2(\xi,\alpha),
\end{eqnarray}
where
\begin{eqnarray}\label{integrals}
F_0(\xi,\alpha) &=& \frac{2}{3} + \tanh\left(\frac{\xi}{\alpha}\right) -
\frac{1}{3}\tanh^3\left(\frac{\xi}{\alpha}\right),\nonumber\\
F_2(\xi,\alpha)
&=&\int_{-\xi/\alpha}^{\infty}\mbox{sech}^4(z)\sin(2z\alpha +
2\xi)dz,
\end{eqnarray}
with the integration variable $z=(x-\xi)/\alpha$. The
Euler-Lagrange equations lead to the following:
\begin{eqnarray}
\alpha_t &=& 4\alpha\beta, \;\;\;\;  \xi_t = 2\kappa,\label{9}\\
\kappa_t &=&
\frac{N}{8\alpha}\frac{\partial}{\partial\xi}\left(\Delta_0 F_0 +
\gamma_1 F_2\right), \\
\beta_t &=& -4\beta^2 + \frac{4}{\pi^2 \alpha^4}
- \frac{N\gamma_0}{\pi^2 \alpha^3}+
\nonumber\\&+&
\frac{3N}{4\pi^2 \alpha}\frac{\partial}{\partial \alpha}
\left(
\frac{\Delta_0 F_0 + \gamma_1 F_2}{\alpha}\right).
\end{eqnarray}
Eliminating the parameter $\beta$ from the equations, we get the
following evolution equation for the width $\alpha$:
\begin{eqnarray}\label{Vareq_a}
\alpha_{tt} &=& \frac{16}{\pi^2 \alpha^3}
- \frac{4N\gamma_0}{\pi^2 \alpha^2}
+
\frac{3N}{\pi^2}\frac{\partial}{\partial\alpha}
\left(\frac{\Delta_0 F_0+\gamma_1F_2}{\alpha}\right).
\end{eqnarray}
This equation can be rewritten as:
\begin{equation}
\alpha_{tt} = -\frac{\partial V_\alpha (\alpha,\xi)}{\partial
\alpha},
\end{equation}
where
\begin{equation}\label{Va}
V_\alpha (\alpha,\xi) = \frac{8}{\pi^2 \alpha^2} -
\frac{N}{\pi^2 \alpha}\left[4\gamma_0 +
3\left(\Delta_0 F_0+\gamma_1 F_2\right)\right].
\end{equation}
In a similar way for the soliton center we get:
\begin{equation}\label{Vareq_x}
\xi_{tt} = -\frac{\partial V_{\xi} (\alpha,\xi)}{\partial\xi},
\end{equation}
where
\begin{equation}\label{Vx}
V_{\xi} (\alpha,\xi) =
- \frac{ N}{4\alpha}\left(\Delta_0F_0 + \gamma_1 F_2\right).
\end{equation}
Typical profiles of effective potentials $V_{\alpha}(\alpha,\xi)$
and $V_{\xi}(\alpha,\xi)$ for the case of narrow solitons are
shown in Fig.~\ref{Veff}. The stationary point given by $\alpha_0$
and $\xi_0$ is obtained in a self-consistent manner, using
Eqs.~(\ref{Va}) and (\ref{Vx}). It should be noted that when the
soliton norm decreases (increase in the width), the amplitude of
the potential $V_{\xi}$ decreases.
Accurate approximative analytical expressions for the effective
potentials $V_{\alpha} (\alpha,\xi)$ and $V_{\xi} (\alpha,\xi)$,
valid in a wide range of variables $\alpha$ and $\xi$, can be
obtained using the asymptotic representation of the integral
(\ref{integrals}) for $F_2(\xi,\alpha)$:
\begin{eqnarray}\label{F2}
F_2(\xi,\alpha) = g(\alpha)\mbox{sech}^4(\xi/\alpha) + \nonumber \\
+\frac{2(1+\alpha^2)\pi \alpha \sin(2\xi)}{3\sinh(\pi
\alpha)}\left( 1 + \tanh(\xi/\alpha) \right),
\end{eqnarray}
where $g(\alpha) \equiv F_2(0,\alpha)$. This expression is
obtained by sewing two different approximations for the integral,
valid at the ranges: $-\infty < \xi \ll -A$, $A \gg 1$, and $a \ll
\xi < \infty$.  The resulting interpolating formulae describes
well the integral $F_2$ in all regions of $\xi$, including the
region $0 < \xi < a$. The dependence of the factor $g(\alpha)$ on
the soliton width is shown in Fig.~\ref{g}.

\section{Results}
All calculations below are performed for a solitary matter wave
with background nonlinearity $\gamma_0 = 1$ and $\Delta_0=0$. In
all the simulations of the soliton transmission and reflection,
the incident wave packets are taken in the standard soliton form,
with the initial parameters given by $A = N/4$, $\alpha_0 = 1/A$,
$\beta = 0$, and $\phi = 0.$ In the starting position, the soliton
is in a homogeneous medium.

\subsection{Stationary soliton}
\begin{figure}[htbp]
\vspace{0.5cm}
\includegraphics[width=8cm,height=6cm,angle=0]{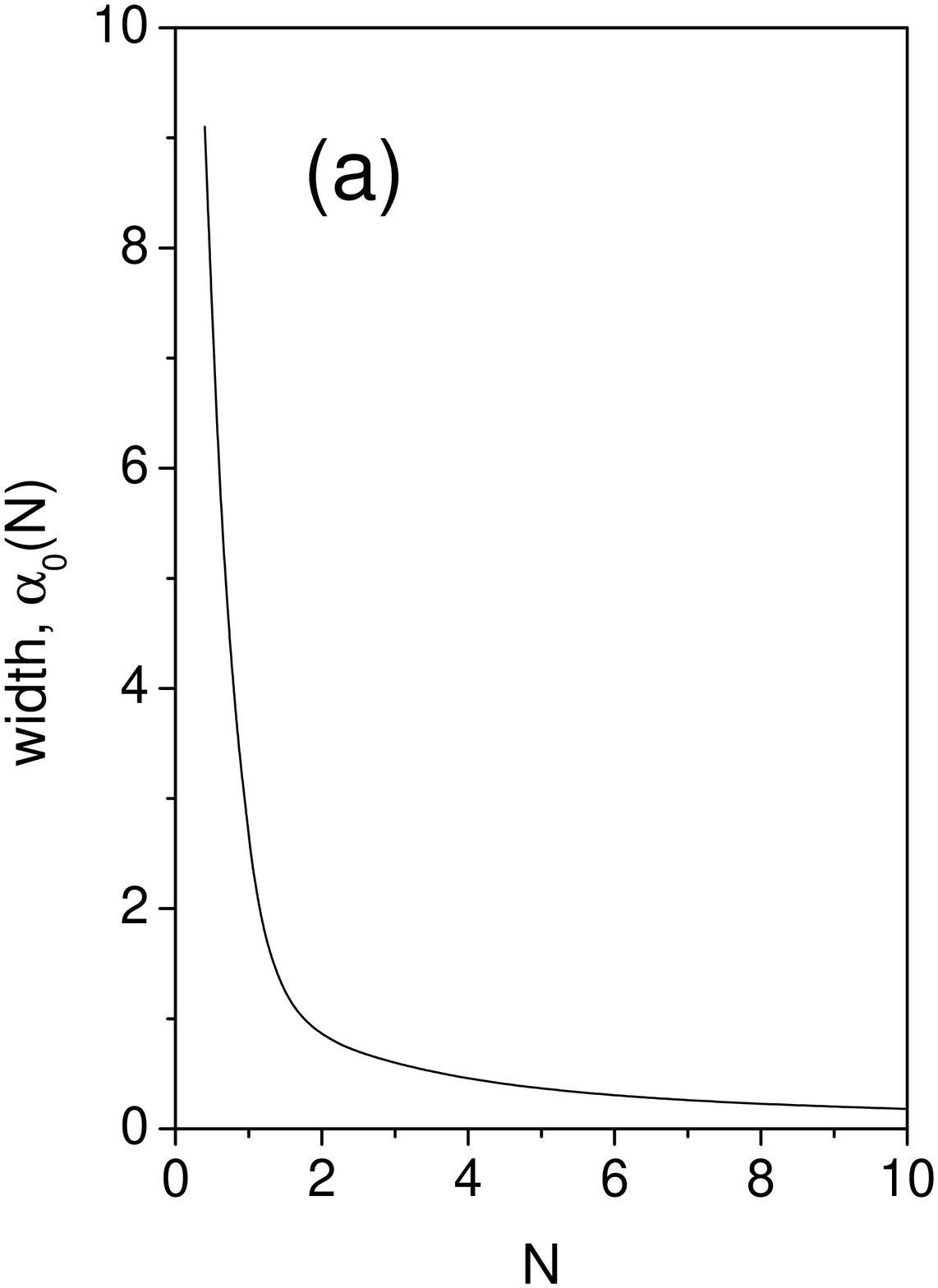}
\hspace{-1cm}
\includegraphics[width=8cm,height=6cm,angle=0]{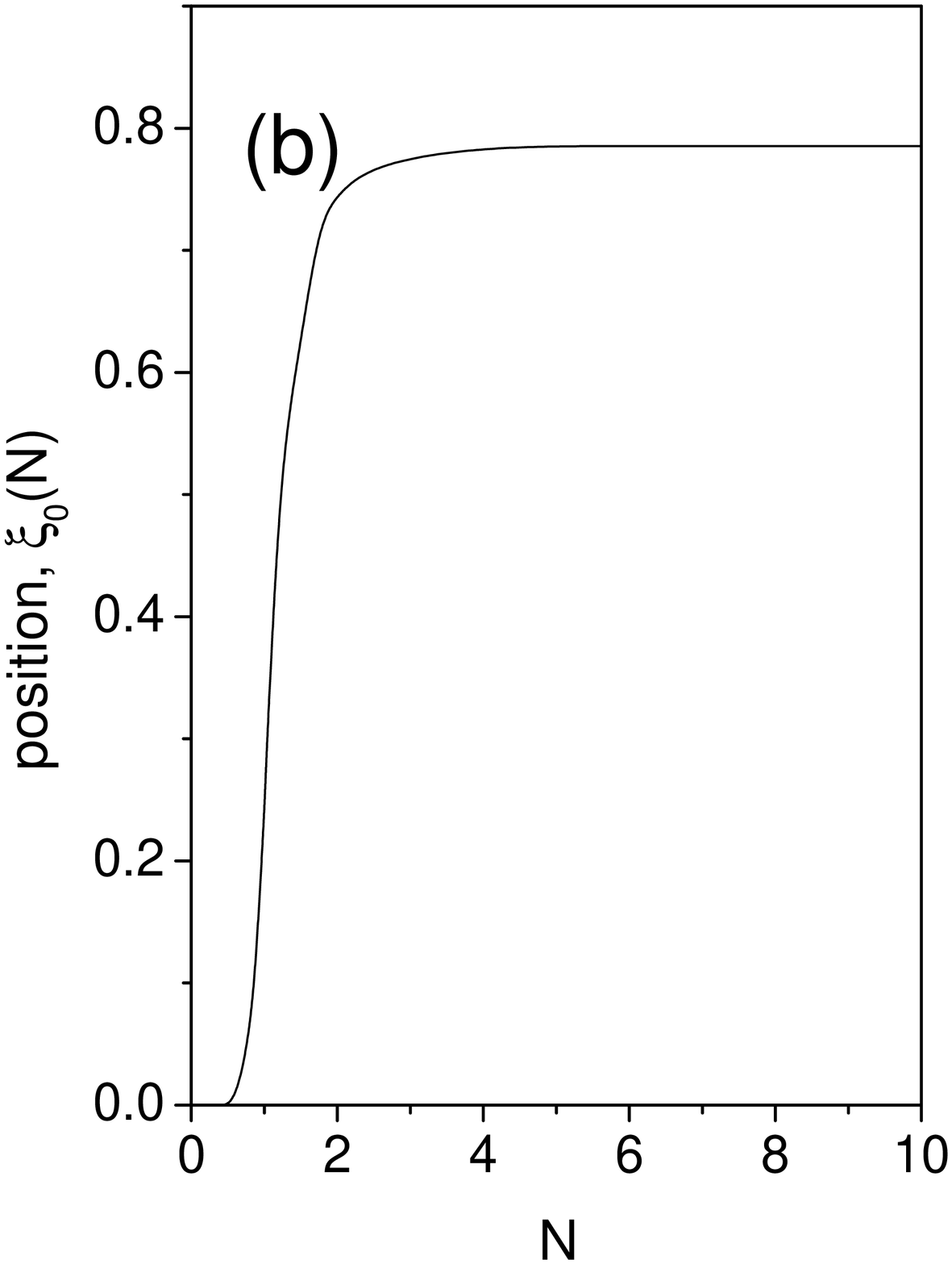}
\caption{Dependences of the width $\alpha_0$ (a) and position
$\xi_0$ (b) of the stationary points on the norm $N$ for the case
$\gamma_1 = 1.2$.} \label{wgN}
\end{figure}

A stationary solution, within a semi-infinite lattice, is given by
expression
\begin{equation}\label{stat}
u_0(x) = \sqrt2 A\mbox{sech} \left(\frac{x-\xi_0}{\alpha_0}\right)
,
\end{equation}
where the stationary values, for the soliton position $\xi$ and
width $\alpha$, are obtained in a self-consistent manner, from the
set of equations
\begin{eqnarray}\label{st:xy}
\displaystyle \left.
\frac{\partial V_{\xi}
(\alpha,\xi)}{\partial\xi}
\right|_{\alpha_0,\xi_0}  &= 0,  \\
\displaystyle \left.
\frac{\partial V_{\alpha}
(\alpha,\xi)}{\partial \alpha} \right|_{\alpha_0,\xi_0} &= 0 .
\label{st:y}
\end{eqnarray}
Fig.~\ref{wgN} depicts stationary values of the soliton width
$\alpha_0$ and its position $\xi_0$ versus the norm $N$.

\begin{figure}[htbp]
\vspace{0.5cm}
\begin{center}
\includegraphics[width=8cm,height=6cm,angle=0]{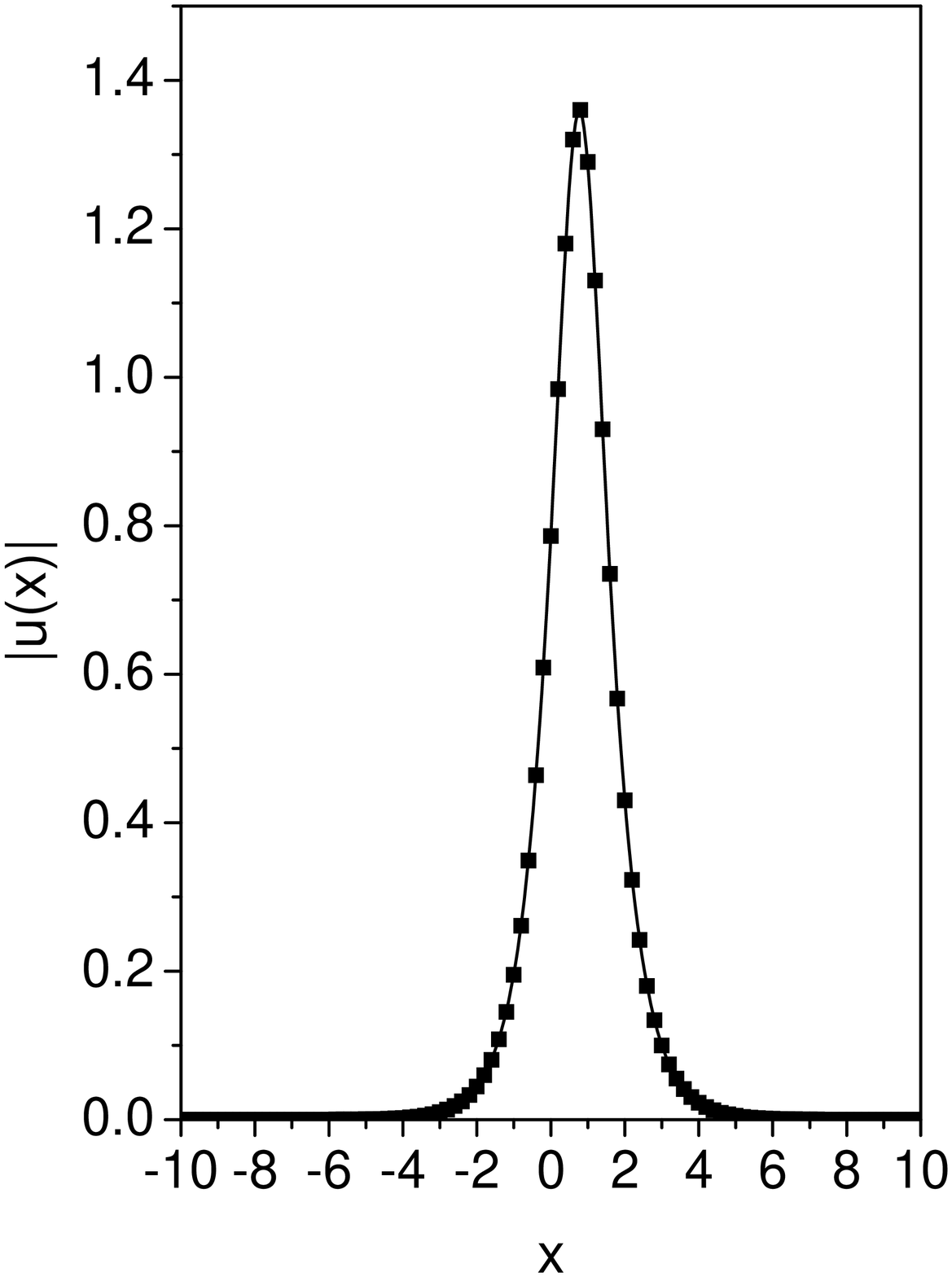}
\end{center}
\caption{ The established stationary wave packet profile $|u(x)|$
at $t=60$. Solid line stands for the full PDE results and squares
are for the single soliton solution given by Eq.~(\ref{trf}). The
parameters are $N=2.5, \ \gamma_1=1.2$. The
stationary point corresponds to $\alpha_0=0.6083$ and
$\xi_0=0.7730$ for the variational approach; and $\alpha_0=0.6293$
and $\xi_0 = 0.7778$ for the full PDE calculation at $t=60$.
} \label{wpack}
\end{figure}

\begin{figure}[htbp]
\vspace{0.5cm}
\begin{center}
\includegraphics[width=8cm,height=6cm,angle=0]{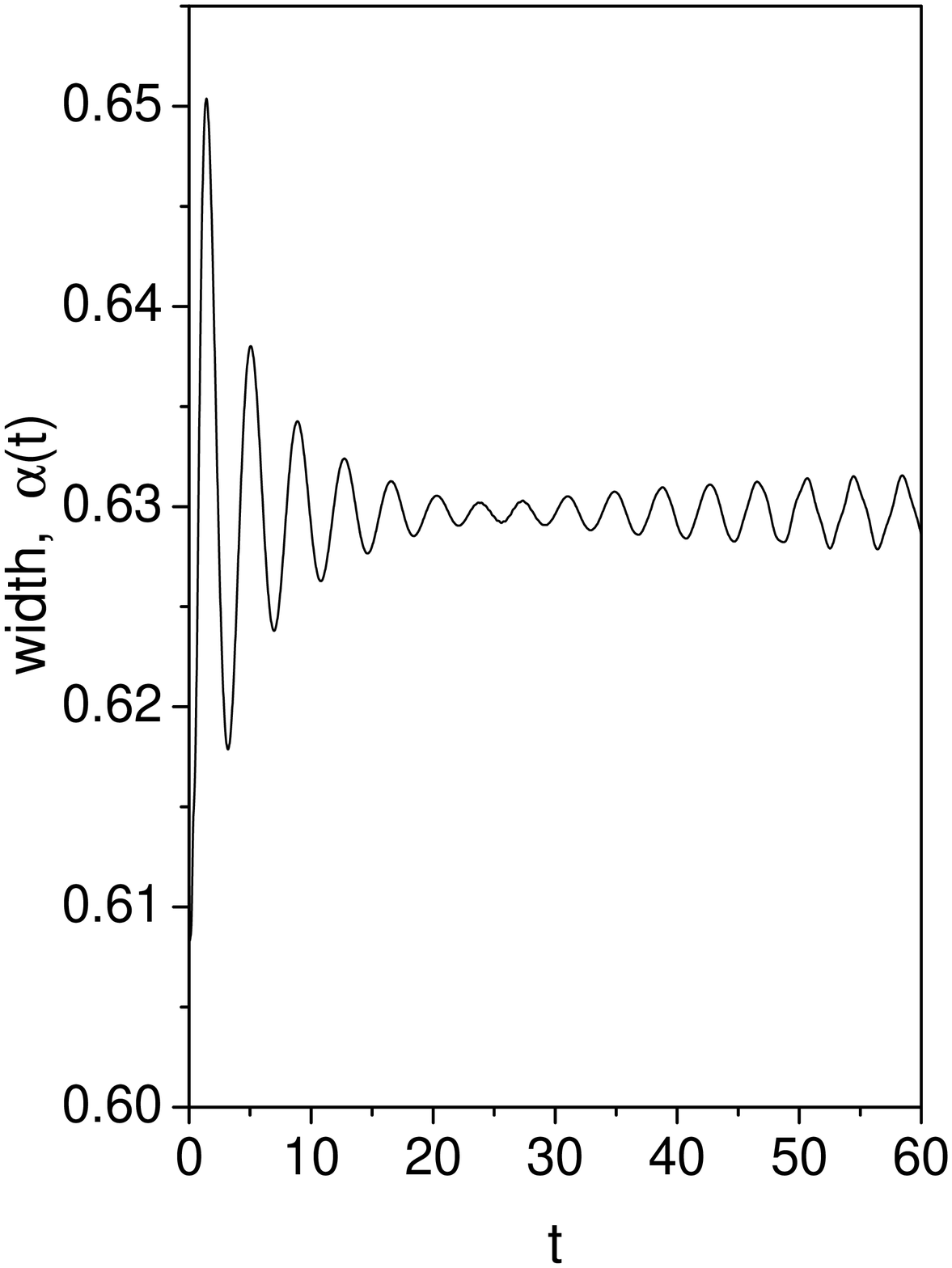}
\end{center}
\caption{Evolution of the wave packet width in the course of
establishment of stationarity when starting from the single
soliton solution given by Eq.~(\ref{trf}). The parameters are
$\mbox{N = 2.5}, \ \alpha_0=0.6083, \ \gamma_1=1.2$.}
\label{wevolv}
\end{figure}

As seen from Fig.~\ref{wgN}(b), the position of the soliton center
shifts from zero to some fixed point value as the number of atoms
increases. In order to check the stationary solution, which was obtained
above, we solve directly the Gross-Pitaeskii equation Eq.~(\ref{eq1}),
with the starting wave packet (\ref{stat}). The results of the full PDE
computation at $t = 60$
are given in Fig.~\ref{wpack}.
The parameters we consider are $N=2.5$, $\gamma_0=1$ and $\gamma_1=1.2$.
In this case, the stationary point obtained from the variational equations
corresponds to $\alpha_0=0.6083$ and $\xi_0=0.7730$.
By starting with this variational point, the results of the full
PDE simulation evolves to $\alpha_0=0.6293$ and $\xi_0 = 0.7778$
at $t=60$.
Establishment of the stationary value of the soliton width in this
case is shown in Fig.~\ref{wevolv} for the evolution of the wave packet.

\begin{figure}[htbp]
\vspace{0.5cm}
\begin{center}
\includegraphics[width=8cm,height=6cm,angle=0]{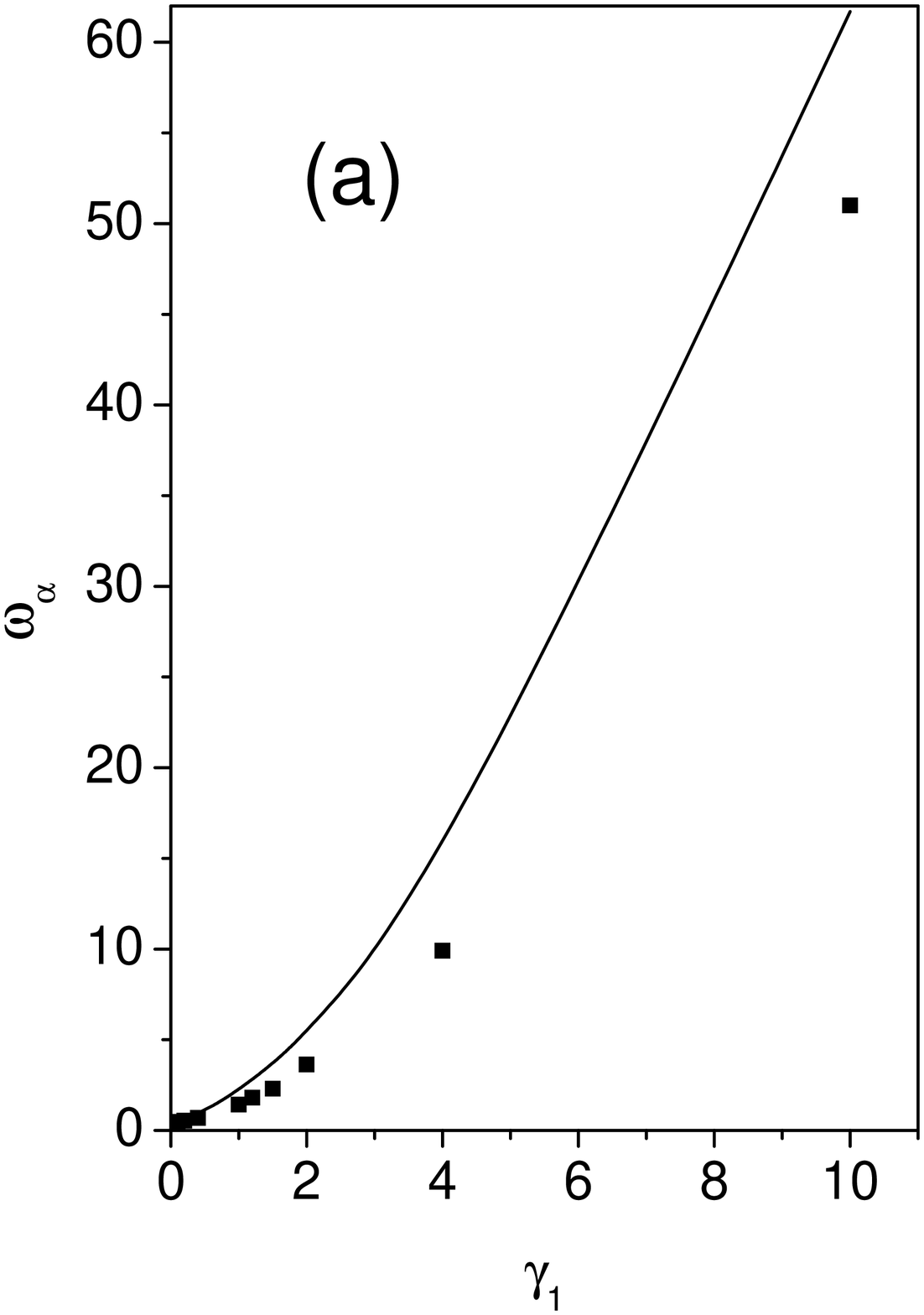}
\hspace{-1cm}
\includegraphics[width=8cm,height=6cm,angle=0]{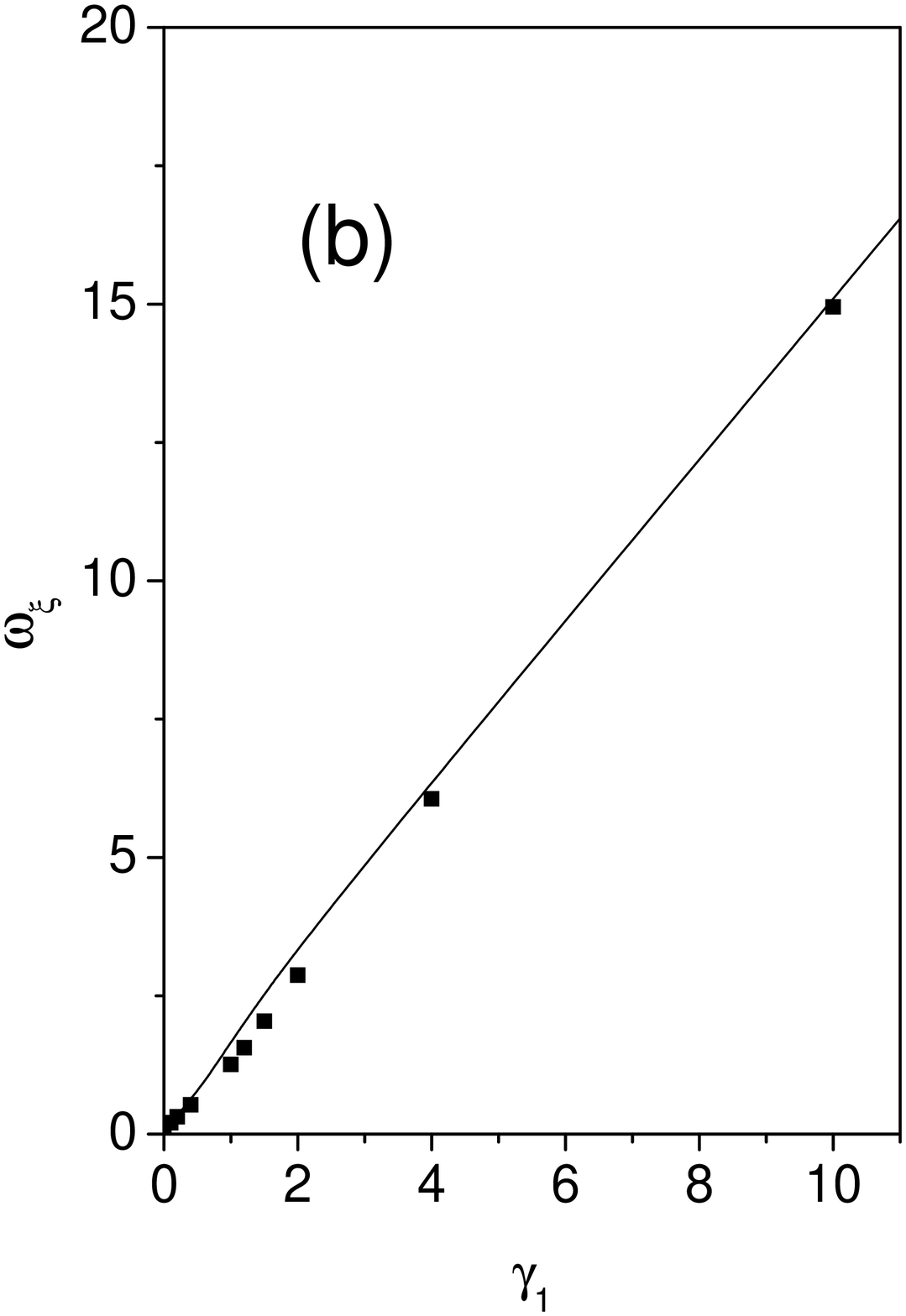}
\end{center}
\caption{Frequencies $\omega_{\alpha}$ and $\omega_{\xi}$ versus
the potential strength $\gamma_1$. With solid lines we show
results obtained with the variational approach. The full PDE
results are shown with squares. The soliton norm is $N = 2.5$.}
\label{Omega-g1}
\end{figure}

We have also calculated the oscillation frequencies at the
stationary point of the potentials $V_{\alpha} (\alpha,\xi)$ and
$V_{\xi}(\alpha,\xi)$ versus the NOL strength $\gamma_1$ for the
case $N = 2.5$ ($\alpha_0 = 1.6$).
The results of our calculations in the frame of the variational
equations (\ref{Vareq_a}), (\ref{Vareq_x}) and PDE simulations are depicted
in Fig.~\ref{Omega-g1}. One can see satisfactory agreement between variational
and PDE results.

\subsection{Reflection and transmission of narrow solitons}

\begin{figure}[htbp]
\vspace{0.5cm}
\begin{center}
\includegraphics[width=8cm,height=6cm,angle=0]{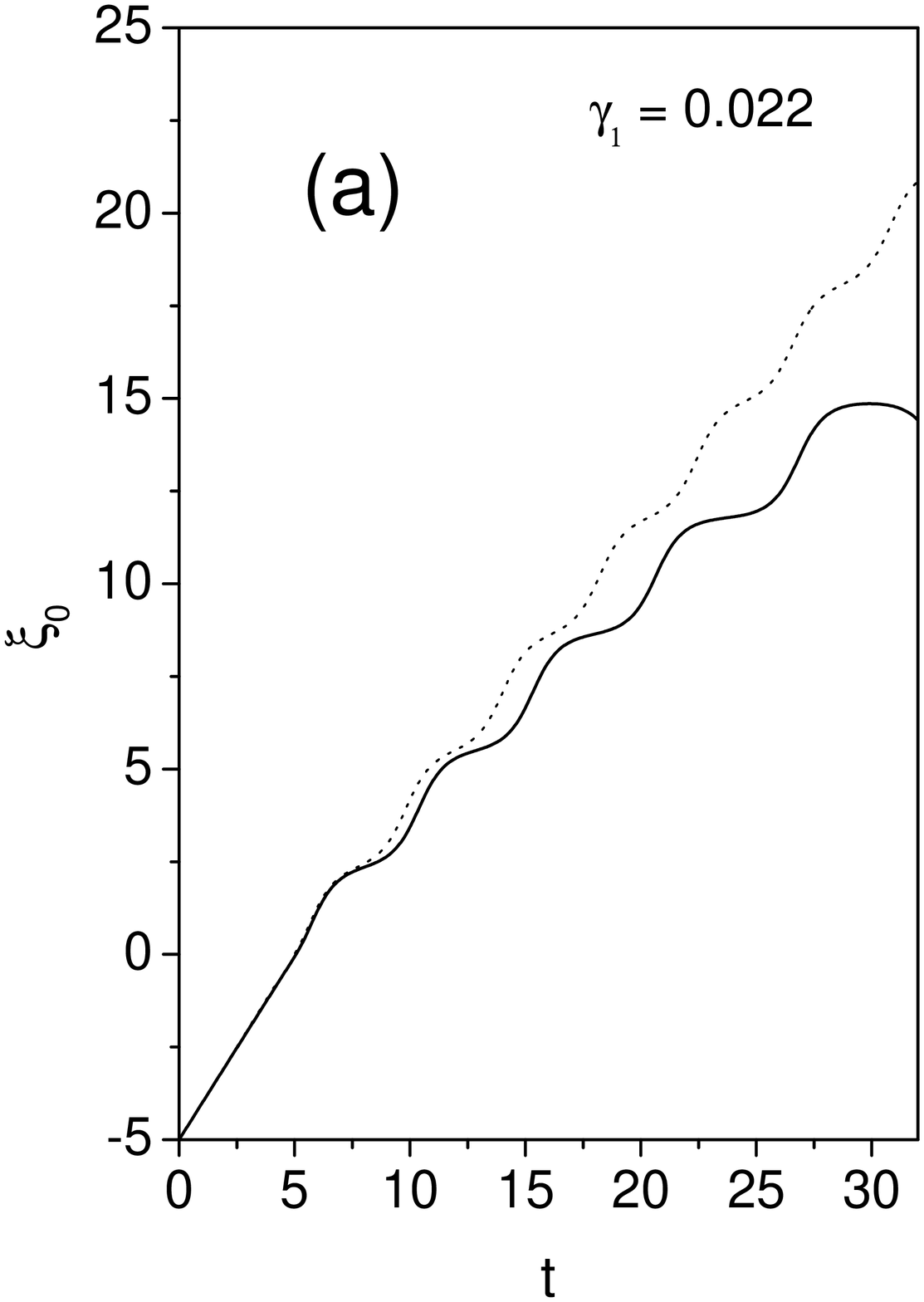}
\hspace{1cm}
\includegraphics[width=8cm,height=6cm,angle=0]{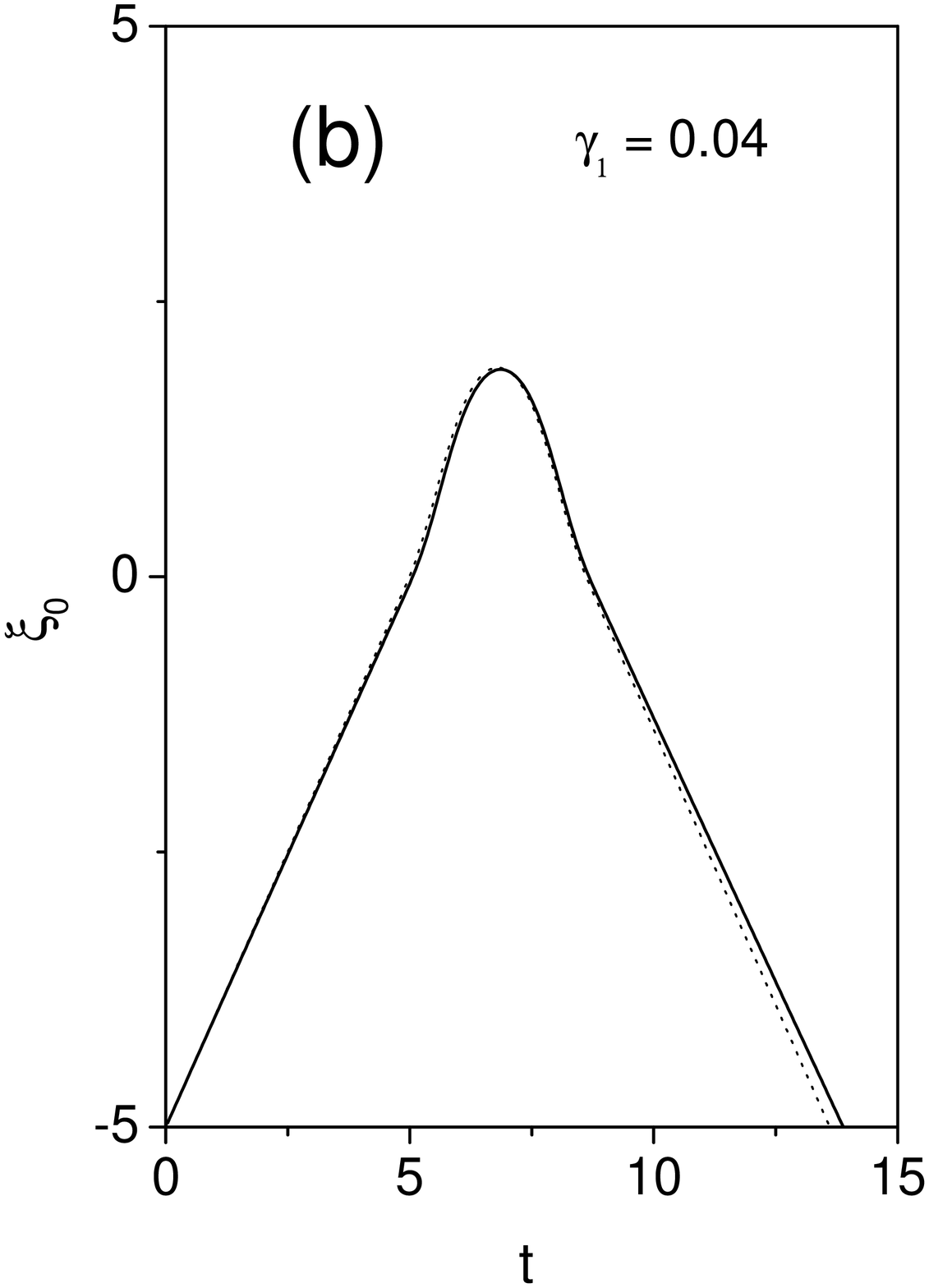}
\end{center}
\caption{Reflection and transmission of a soliton  below (a) and
above (b) the threshold value of $\gamma_1 = 0.023$. Solid lines
stand for PDE simulations and dotted lines for variational ODE
calculations. The parameters of the soliton are $\alpha_0 = 0.25$, $N
= 16$, $v = 1$.} \label{Refl}
\end{figure}

Reflection and transmission of solitons at the interface between a
homogeneous medium and the optical lattice can be described by
Eq.~(\ref{Vareq_x}) considering the soliton as a particle
travelling under the effective potential $V_{\xi}(\alpha,\xi)$.
Then, the condition of reflection or transmission is determined by
the potential barrier height (see Fig.~\ref{Veff}). In
Fig.~\ref{Refl} we present simulations for transmission and
reflection of the soliton at the interface, above and below the
threshold value of the nonlinearity strength $\gamma_1$.
The results obtained with ODE calculations, based on
Eq.~(\ref{Vareq_x}), are shown with dotted lines. The solid lines
correspond to PDE simulations, obtained from the GP equation
(\ref{eq1}). The soliton parameters are $\alpha_0 = 0.25$, $N = 16$
and $v = 1$, where $v \equiv \xi_t = 2\kappa$ is the soliton velocity
(see Eq.~(\ref{9})).
The potential barrier heights,
$\gamma_1 = 0.022$, indicated in Fig.~\ref{Refl}(a) and $\gamma_1
= 0.04$, indicated in Fig.~\ref{Refl}(b), are respectively below
and above the kinetic energy of the soliton, $E_{kin} = v^2/2$.
As seen, the transmission and reflection conditions obtained in this case are
in a good agreement with the PDE simulations of Eq.~(\ref{eq1}). We also
observe that the true soliton dynamics deviates from the
analytical prediction for quite large times. At some depth of the
soliton penetration into the optical lattice, the soliton can be
trapped due to the radiation effects.

\begin{figure}[htbp]
\includegraphics[width=8cm,height=6cm,angle=0]{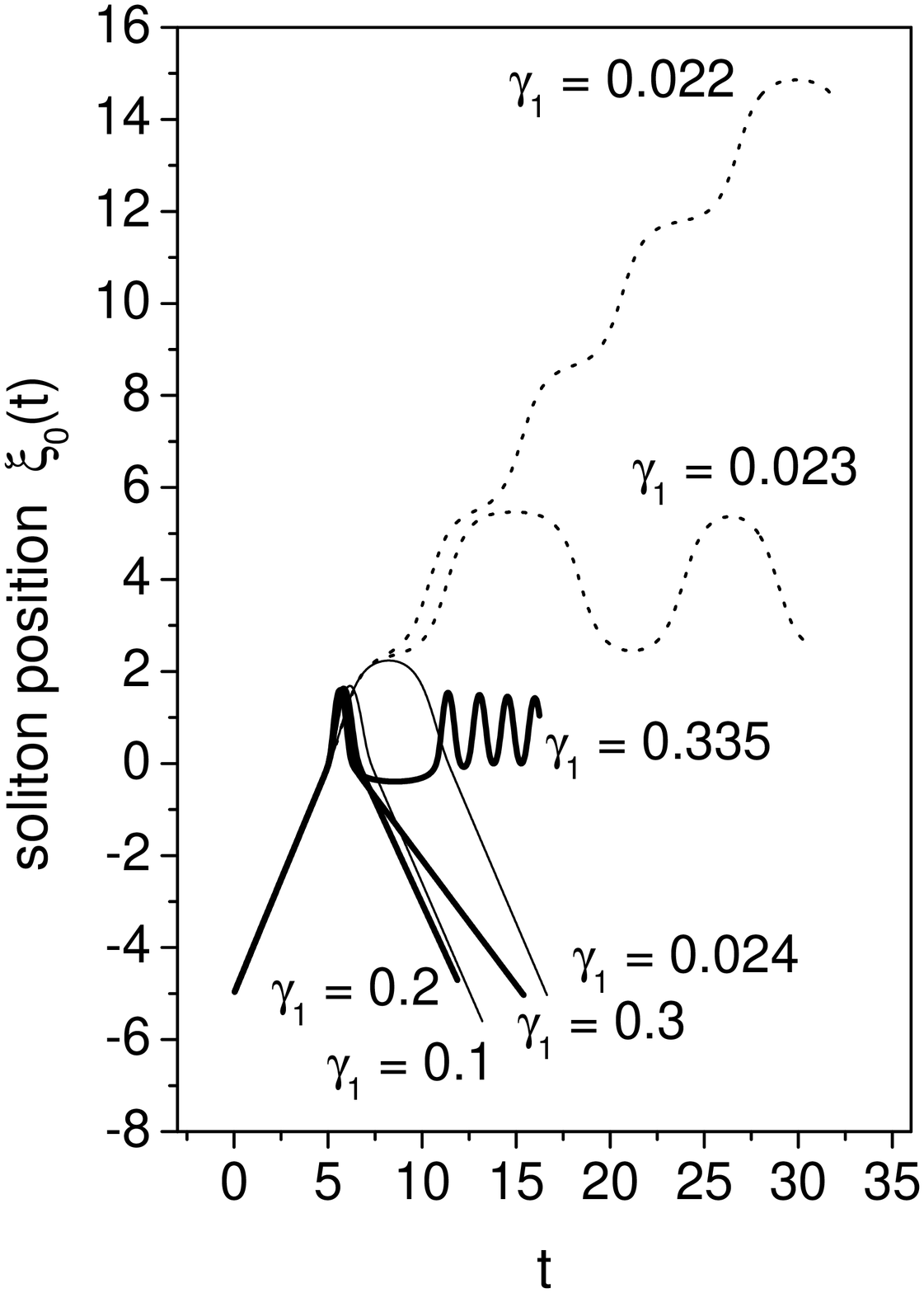}
\caption{Travelling of the position of a narrow soliton at
different strengths of the periodic potential. Processes of
transmission, reflection and trapping are depicted. The soliton
parameters are $\alpha_0=0.25$, $N=16$ and $v=1$.}
\label{Rf-vs-g1}
\end{figure}

\begin{figure}[htbp]
\begin{center}
\includegraphics[width=8cm,height=6cm,angle=0]{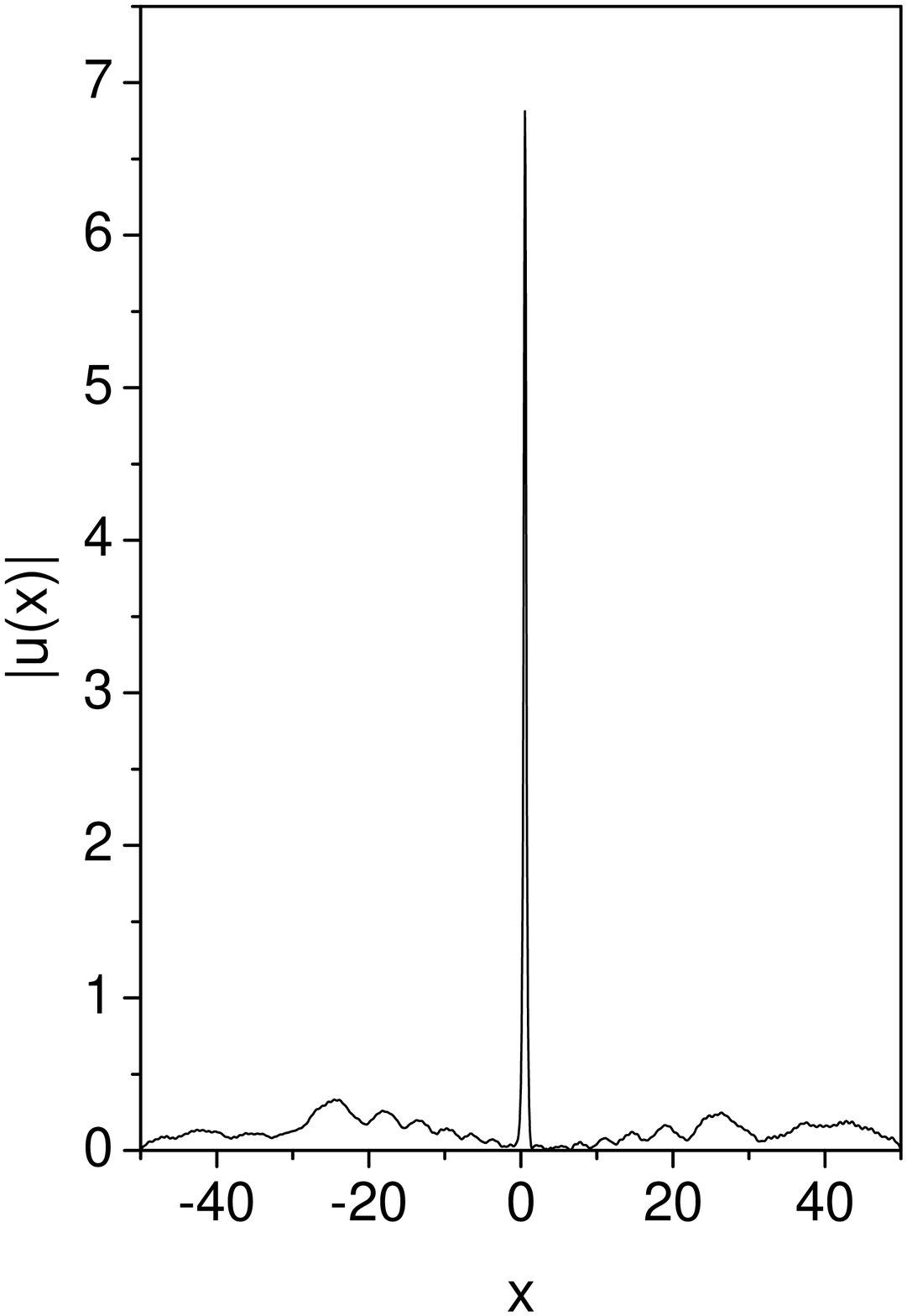}
\end{center}
\caption{Trapped soliton at $t=10$. The simulation is started from
the single soliton solution given by Eq.~(\ref{trf}) moving with
velocity $v=1$. Its initial position is $\xi_0 = -5.$ Other
parameters are $N=16$, $\gamma_0=1$, $\gamma_1=0.4$, and
$\alpha_0=0.25$.} \label{trapping}
\end{figure}

\begin{figure}[htbp]
\vspace{0.5cm}
\begin{center}
\includegraphics[width=8cm,height=6cm,angle=0]{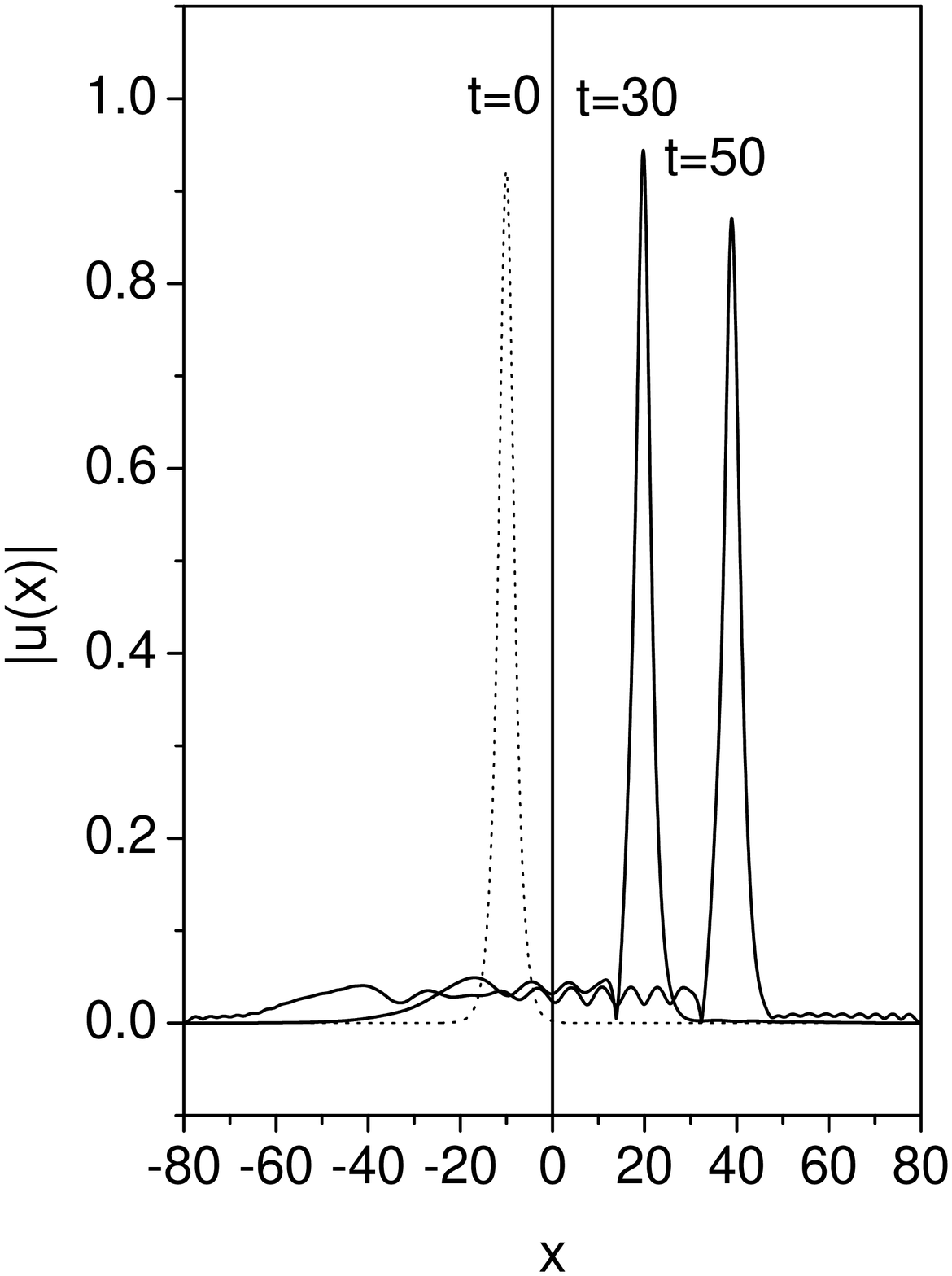}
\end{center}
\caption{Transmission of the narrow wave packet when starting from
the single soliton solution with the parameters $N = 2.721, \
\alpha_0=1.47, \ \xi_0 = -10, \ v = 1$. The nonlinear optical
lattice strength is $\gamma_1=0.1$.} \label{transm}
\end{figure}

To study numerically the travelling of a narrow soliton in a wide
range of strengths of the optical lattice (0.022$<\gamma_1<0.4$)
we carried out corresponding PDE simulations presented in
Fig.~(\ref{Rf-vs-g1}). The soliton parameters were $\alpha= 0.25$,
$N = 16$, and $v = 1$.
One can observe two regions of the optical lattice strength
$\gamma_1$ which provide trapping of the soliton: $\gamma_1 \leq
0.023$ and $\gamma_1 \geq 0.335$.
The first region ($\gamma_1 \leq 0.023$) corresponds to the
soliton motion above the barrier. The trapping in this case is
caused by an unavoidable radiation, which decreases the soliton
kinetic energy, in the course of its motion in the optical
lattice.
The cause of the soliton trapping in the second case ($\gamma_1
\geq 0.335$) can be explained by rearrangements of
the wave packet due to deepening of the effective potential well,
accompanied by strong radiation that results in transformation of
the incident {\it moving} soliton to the {\it stationary} one.
Typical profile of the trapped
soliton in this case is depicted in Fig.~(\ref{trapping}).
The evolution of the soliton wave packet profile in the course of
transmission of the optical lattice is presented in Fig.~(\ref{transm}).
As the soliton penetrates in the optical lattice, we observe that
the amplitude of the transmitted soliton decreases due to
noticeable radiation.

\subsection{Broad soliton}

\begin{figure}[htbp]
\vspace{0.5cm}
\begin{center}
\includegraphics[width=8cm,height=6cm,angle=0]{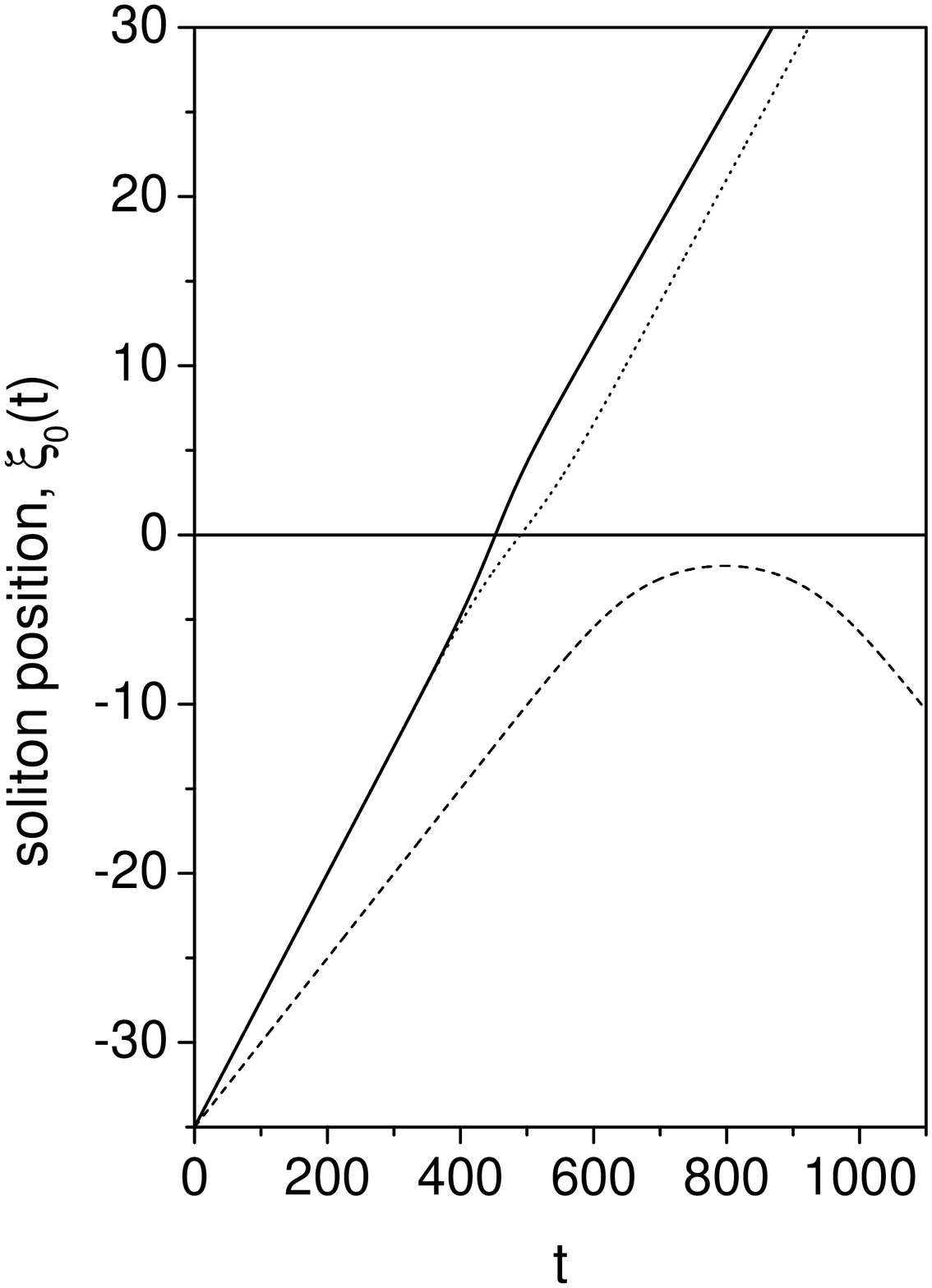}
\end{center}
\caption{Numerical simulations of reflection and transmission of a
broad soliton with the parameters $N = 0.6$ and $\alpha_0 = 6.667$.
The solid line stands for the case $\gamma_1 > 0$, when the
effective potential $V_{\xi}(\alpha,\xi)$ is a potential well.
Dotted and dashed lines are for the case $\gamma_1 < 0$, when the
effective potential is a barrier
for an incident soliton.
The soliton velocities $v = 2\kappa = 0.075$ (dotted line) and
$v = 2\kappa = 0.05$ (dashed line) correspond to the soliton
kinetic energies upper and below the potential barrier, respectively.}
\label{broadRef_Tr}
\end{figure}

Let us describe the dynamics of the broad soliton by
Eqs.~(\ref{Vareq_x}) and (\ref{Vareq_a}). The VA is works well for
the  soliton propagation with the width less or of the  order of
the lattice period. The validity of VA for the dynamics of a broad
(with respect to the lattice period) soliton  should be checked by
 direct numerical simulations. As a rule  one can  expect a good
agreement for the VA if radiation effects in  propagation of the
soliton  in periodically modulated media are
small~\cite{malomed02}.  As shown in Ref.~\cite{AG05}, radiative
effects at motion in the media with spatially periodic
nonlinearity are small at the propagation of soliton with  small
velocity.  The approximated expressions for the effective
potentials $V_{\alpha}(\alpha,\xi)$ and $V_{\xi}(\alpha,\xi)$ can
be simplified when considering large $\alpha$. So, as the soliton
width grows to large $\alpha$, the second term of Eq.~(\ref{F2})
can be neglected. By also imposing
$\Delta_0 = 0$, the effective potential $V_{\xi}(\alpha,\xi)$
takes the form
\begin{eqnarray}\label{brVx}
V_{\xi}(\alpha,\xi) = -\frac{N \gamma_1}{4}
\frac{g(\alpha)}{\alpha}\mbox{sech}^4(\xi/\alpha).
\end{eqnarray}
For $\gamma_1>0$, the potential $V_{\xi}(\alpha,\xi)$ is a
potential well and, for $\gamma_1<0$, a potential barrier. It
means that the reflection of the soliton becomes only possible
provided that $\gamma_1 < 0$. The sign of $\gamma_1$ is defined by
the phase $\delta$ of the periodic modulation of the nonlinearity
$\sim \sin(2x + \delta)$. Thus, with the variation of such phase
one can switch the matter-wave soliton from a transmission regime
to a reflection one. The transmission (reflection) of the soliton
occurs when the soliton kinetic energy is greater (smaller) than
the potential barrier height. The threshold kinetic energy,
$E_{cr}$, is given by
\begin{equation}\label{Ekin}
E_{cr} = \frac{v^2}{2} = |V_{\xi}(\alpha,\xi)|_{max} =
N |\gamma_1| \frac{g(\alpha)}{4\alpha}.
\end{equation}
Fig.~(\ref{broadRef_Tr}) depicts transmission of a broad soliton
for the case $\gamma_1 > 0$, when the effective potential
$V_{\xi}(\alpha,\xi)$ is a potential well; and reflection and
transmission of the soliton for the case $\gamma_1 < 0$, when the
effective potential is a barrier. As seen, the conditions of
transmission and reflection are well described by
Eq.~(\ref{Ekin}).
It should be noted that trapping of the soliton is not observed
since the effective potential $V_{\xi}(\alpha,\xi)$ is a
short-range one, such that, far from the interface, motion of the
soliton can be considered as free.

 Let us discuss the possible effects which can be predicted in
experiments with BEC in a cigar-type trap. As an example, we can
consider the $^7$Li condensate in the elongated trap with the
transverse frequency $\omega_{\perp}\approx 2\pi \times 10^3$Hz
and the longitudinal frequency $\omega_x \approx $ few Hz. The
density is $n \approx 10^9 m^{-1}$. The healing length and speed
of sound are $\xi \approx 2\mu$m and $c \approx 5$mm/s. In a
typical experiment~\cite{Hulet} we could consider a soliton with
about $10^3$ atoms and width $\approx 2\xi \approx 4\mu$m.  In
experiments we can vary the scattering length by using the
Feshbach resonance method by varying in space the external
magnetic field $B(x)$ near the resonant value $B_c$, such that
$$
a_s(x)=a_b\left( 1 - \frac{\Delta}{B_c - B(x)}\right),
$$
where $a_b$ is the background scattering length and $\Delta $ is
the resonance width~\cite{Pitaevskii}. Other way to vary the
scattering length is the using of the optically induced Feshbach
resonances~\cite{Kagan,NOL1,NOL2}. Typical values of variations
 of the external field B are in the interval
$(0-200)\mu$m with the period $10\mu$m around the value 352 G,
where the scattering length has the minimal value $\approx
-0.23$nm.

\section{Conclusion}
We have investigated reflection, transmission and trapping of a
matter-wave soliton, which is propagating through the interface
between a nonlinear uniform media and a nonlinear optical lattice.
We study analytically  two different limits for broad and narrow
soliton dynamics. In both the cases we obtain the characteristics
of a soliton trapped by an interface, corresponding to a localized
nonlinear surface wave. We derive the effective potentials for the
soliton center-of-mass and the width induced by the joint action
of the interface and the media periodic nonlinearity. We obtain
the parameters of a localized nonlinear surface state,
corresponding to a soliton trapped near the interface. Near the
stationary point of effective potentials Eq.~(\ref{Va}) and
Eq.~(\ref{Vx}), we have calculated frequencies of oscillations of
the trapped soliton center and its width. It was also obtained the
threshold value of the NOL strength $\gamma_1$, separating the
transmission and reflection regimes for incident solitons. The
predicted surface soliton states can be observed in experiments
with BEC in optically induced NOL and in nonlinear optical systems
with periodically modulated Kerr nonlinearity.

\section*{Acknowledgements}
 Authors are grateful to B.B. Baizakov and E.N. Tsoy for useful
 discussions. F.Kh.A. is grateful to the grant SAGA
Fund 77 by MOSTI for a partial support. M.B. and L.T. thank
Funda\c c\~ao de Amparo \'a Pesquisa do Estado de S\~ao Paulo
(FAPESP) and Conselho Nacional de Desenvolvimento Cient\'\i fico e
Tecnol\'ogico (CNPq) for partial support.


\begin{thebibliography}{99}
\bibitem{ANM}
A.B. Aceves, J. Moloney, and A.C. Newell, Phys. Rev. A. {\bf 39},
1809 (1989).

\bibitem{KKC}
Y.S. Kivshar, A.M. Kosevich, and O.A. Chubykalo, Phys. Rev. A {\bf
41}, 1677 (1990).

\bibitem{Kartashov06}
Y.V.  Kartashov, V.A. Vysloukh, and L. Torner, Opt. Express {\bf
14}, 1576 (2006).

\bibitem{Kagan}
P.O. Fedichev, Yu. Kagan, G.V. Schlyapnikov, and J.T.M. Walraven,
Phys. Rev. Lett. {\bf 77}, 2913  (1996).

\bibitem{NOL1}
 F. K. Fatemi, K. M. Jones, and P. D. Lett,
 Phys. Rev. Lett. {\bf 85}, 4462 (2000).

\bibitem{NOL2}
 M. Theis, G. Thalhammaer, K. Winkler, M. Hellwig, G. Ruff, R.
Grimm, and J.H. Denschlag, Phys. Rev. Lett. {\bf 93}, 123001
(2004).

\bibitem{AS03}
F.Kh. Abdullaev and  M. Salerno, J. Phys. B {\bf 36}, 2851 (2003).

\bibitem{SM06}
H. Sakaguchi and B.A. Malomed, Phys. Rev.E {\bf 72}, 046610
(2005).

\bibitem{AG05}
F.Kh. Abdullaev and J. Garnier,  Phys. Rev. A {\bf 72}, 061605(R)
(2005).

\bibitem{kevrekidis}
P.G. Kevrekidis et al. Phys. Rev. A {\bf 72}, 033614 (2005).

\bibitem{Garcia07}
J. Belmonte-Beitia, V.M. Perez-Garcia, V. Vekslerchik, and P.
Torres,  Phys. Rev. Lett. {\bf 98}, 064102 (2007).

\bibitem{Abd08}
F.Kh. Abdullaev, A.Gammal, H.L.F. da Luz, and L. Tomio, Phys. Rev.
A {\bf 76}, 043611 (2007).

\bibitem{Kart}
Y.V. Kartashov, V.A. Vysloukh, L. Torner,
eprint arXiv:0808.3357v1.

\bibitem{Blud}
Yu. V. Bludov and V. V. Konotop,
Phys. Rev. A {\bf 74}, 043616 (2006).

\bibitem{Konotop}
M. Salerno, V. V. Konotop, and Yu. V. Bludov,
Phys. Rev. Lett. {\bf 101}, 030405 (2008).

\bibitem{AGST} F.Kh.  Abdullaev, A. Gammal, M. Salerno, and L. Tomio,
Phys. Rev. A {\bf 77}, 023615 (2008).

\bibitem{Dong}
G. Dong and B.Hu, Phys. Rev.A {\bf 75}, 013625 (2007).

\bibitem{ijmpe} F.Kh. Abdullaev, A. Gammal, A.M. Kamchatnov, and L. Tomio,
Int. J. of Mod. Phys. B {\bf 19}, 3415 (2005).

\bibitem{AAG}
F.Kh. Abdullaev, A.A. Abdumalikov, and R.M.  Galimzyanov, Phys.
Lett. A {\bf 367}, 149 (2007).

\bibitem{fibich}
G. Fibich, Y. Sivan, and M. Weinstein, Physica D {\bf 217}, 31
(2006); Y. Sivan, G. Fibich and M. Weinstein, Phys. Rev. Lett.
{\bf 97}, 193902 (2006).

\bibitem{Garanovich}
I.L. Garanovich, A.A. Sukhorukov, and Y.S. Kivshar,
Phys. Rev. Lett. {\bf 100}, 203904 (2008).

\bibitem{Kominis}
Y. Kominis and K. Hizanidis, Phys. Rev. Lett. {\bf 102}, 133903
(2009).

\bibitem{Kartashov3}
Y.V. Kartashov, V.A. Vysloukh, A. Szameit, F. Dreisow, M.
Heinrich, S. Nolte, A. Tunnrmann, T. Pertsch, and L. Torner, Opt.
Lett. {\bf 33}, 1120 (2008).

\bibitem{gtf}
A. Gammal, L. Tomio, and T. Frederico, Phys. Rev. A {\bf 66}, 043619 (2002);
A. Gammal, T. Frederico, and L. Tomio, Phys. Rev. A {\bf 64}, 055602 (2001).

\bibitem{anderson}
D. Anderson, Phys. Rev. A {\bf 27}, 3135 (1983).

\bibitem{malomed02}
B.A. Malomed, Progress in Optics {\bf 43},69 (2002).

\bibitem{Hulet}
K.E. Strecker, G.B. Patridge, A.G. Truscott, R.G. Hulet, Nature
{\bf 417}, 150(2002); L. Khaykovich, F. Schreck. G. Ferrari, T.
Bourdel, J. Cubizolles, L.D. Carr, Y.Castin, C. Salomon, Science
{\bf 256}, 1290 (2002).

\bibitem{Pitaevskii}
L.P. Pitaevskii and S. Stringari, {\it Bose-Einstein
condensation}, Oxford Univ.Press (2004).


\end{thebibliography}
\end{document}